# Particle Tracking of Microelectromechanical System Performance and Reliability

Craig R. Copeland, Craig D. McGray, Jon Geist, and Samuel M. Stavis

*Abstract*—Microelectromechanical systems (MEMS) that require contact of moving parts to implement complex functions exhibit limits to their performance and reliability. Here, we advance our particle tracking method to measure MEMS motion *in operando* at nanometer, microradian, and millisecond scales. We test a torsional ratcheting actuator and observe dynamic behavior ranging from nearly perfect repeatability, to transient feedback and stiction, to terminal failure. This new measurement capability will help to understand and improve MEMS motion.

## I. Introduction

MICROELECTROMECHANICAL systems (MEMS) that output motion with multiple degrees of freedom or multiradian rotations often require mechanisms such as bearings, linkages, ratchets, and gears. However, kinematic assemblies of parts in sliding contact can be subject to coupling interactions that limit their performance and reliability. Understanding and mitigating these interactions is a topic of intense interest [1]–[5]. In particular, surface forces cause dynamic changes in MEMS behavior [6]–[8]. Friction can result in adhesion and wear of parts, and mechanical feedback can degrade kinematic precision. These interactions are critical in systems with multifunctional parts where motion from one function affects another function.

Measuring the motion and interaction of parts is essential to understand and improve MEMS performance and reliability, and emerging methods of superresolution microscopy provide new capabilities to do so [4], [9]–[11]. Recently, we developed a particle tracking method to resolve MEMS motion at nanometer and microradian scales. Our method is broadly applicable to diverse devices, requires only simple instrumentation, and is highly customizable within the measurement framework of optical microscopy [12]. In this Letter, we scale up the speed of our measurements by two orders of magnitude, enabling millisecond resolution and accelerating repetitive tests. Importantly, our method can resolve the fine details of transient and aperiodic motion leading up to and during system failure.

We apply our method to test a torsional ratcheting actuator [13] (Fig. 1(a)). Through its intricate assembly of compliant and rigid

This work was supported by the National Institute of Standards and Technol-ogy (NIST) Innovations in Measurement Science Program. The work of C. R. Copeland was supported by the Cooperative Research Agree-ment between the University of Maryland and the NIST Center for Nanoscale Science and Technology through the University of Maryland, under Grant 70ANB10H193.
*(Corresponding author: Samuel M. Stavis.)*
The authors are with the National Institute of Standards and Technology, Gaithersburg, MD 20899 USA (e-mail: samuel.stavis@nist.gov).

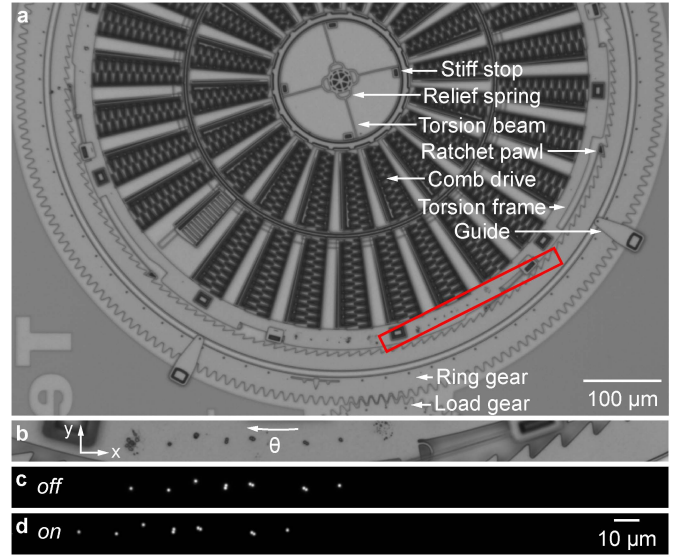

Fig. 1. Experimental overview. (a) Brightfield optical micrograph showing a torsional ratcheting actuator at inspection magnification. The red rectangle denotes the region of interest in (b-d). (b) Brightfield optical micrograph showing the region of interest at experimental magnification. (c) Fluorescence optical micrograph showing the same region as (b). Fluorescent particles label the torsion frame. (d) Fluorescence optical micrograph showing a displacement of approximately 15 $\mu$m and a rotation of approximately 48 $\mu$rad with respect to (c). Supplementary Video 1 shows representative motion from Fig. 2.

mechanisms, this remarkable system operates in an open loop and requires only simple input signals of tens of volts to output forces of several micronewtons, displacements of tens of micrometers, and rotations of tens of milliradians for single motion cycles [13]. This system is also one of the more reliable MEMS that transfer motion through parts in sliding contact [14]. Still, the resulting interactions lead to variable performance and untimely failure, and are not fully understood, which has discouraged the application of such technology. Our method demonstrates the potential to better measure MEMS motion at the limits of performance and reliability, and inform the future development of MEMS to realize their full potential.

## II. Materials and Methods

### A. Test System

Application of an input voltage $V_{\text{in}}$ actuates an electrostatic comb array, rotating a torsion frame by $\Delta\theta$ (Fig. 1(b)). Torsion beams and relief springs link the torsion frame to the substrate and apply a restoring torque, rotating the torsion frame back to $\theta \approx 0$ upon $V_{\text{in}}$ cessation. Thus, application of a periodic $V_{\text{in}}$ results in oscillation of the torsion frame, which couples by ratchet pawls to a ring gear, rotating the ring gear by approximately 1 mrad with each oscillation. The ring gear meshes with and rotates a load gear.

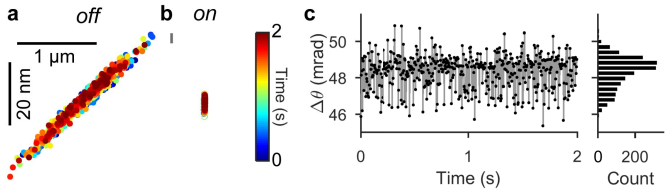

Fig. 2. Normal operation. (a) Scatter plot showing *off* positions. (b) Scatter plot showing *on* positions. The mean displacement between the *off* and *on* positions is 15.52 $\mu$m. (c) Time series and histogram showing rotations.

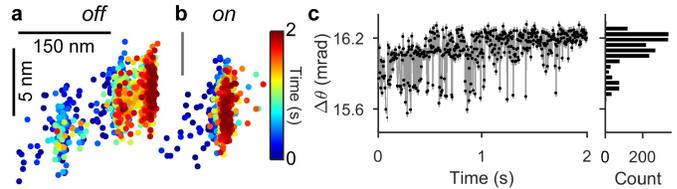

Fig. 3. Feedback effects. (a) Scatter plot showing *off* positions. (b) Scatter plot showing *on* positions. The mean displacement between the *off* and *on* positions is 5.19 $\mu$m. (c) Time series and histogram showing rotations.

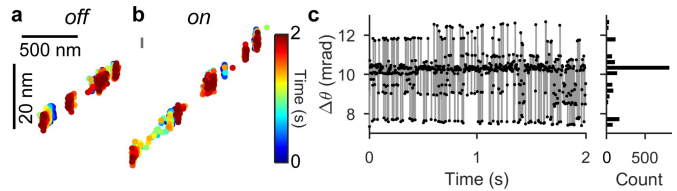

Fig. 4. Stiction effects. (a) Scatter plot showing *off* positions. (b) Scatter plot showing *on* positions. The mean displacement between the *off* and *on* positions is 3.22 $\mu$m. (c) Time series and histogram showing rotations.

### B. Measurement Method

We measure planar motion by particle tracking [4], [9], [10]. Briefly, we deposit fluorescent particles onto the torsion frame and synchronize imaging with a square wave of variable $V_{in}$ to drive the quasistatic motion of the torsion frame between *off* and *on* states (Fig. 1(c) and (d)) [4]. We localize single particles and determine the rigid transformations that map the particle positions between consecutive images. In this way, we measure the centroid positions in x and y and $\Delta\theta$ of the torsion frame over 2000 motion cycles per experiment. We calibrate our microscope to ensure accurate localization [15], eliminating errors due to nonuniform magnification. We also correct errors due to rotation of the torsion frame out of the imaging plane [4]. With respect to our previous work, we trade off some spatial range for temporal resolution [12], by decreasing the readout height of our complementary metal–oxide–semiconductor (CMOS) camera to 128 pixels, and increasing the readout rate to 1 kHz. To maintain uncertainties at the nanometer and microradian scales, we increase both the illumination irradiance and particle diameter, boosting the mean flux of signal photons by a factor of $2.9 \times 10^2$. We determine uncertainties of 1.6 nm for position in x and y and 23 $\mu$rad for $\theta$ as the standard deviations of these values from a nominally static system over 2000 images. Uncertainties of displacement and $\Delta\theta$ are larger by a factor of $\sqrt{2}$.

## III. RESULTS AND DISCUSSION

For brevity, we omit a global analysis of $\theta(V_{in})$ [13]. We note that particle tracking and empirical modeling can accurately describe the system response in the presence of experimental defects, which theoretical models generally do not account for [4]. Instead, we focus on the intriguing details of the results from a series of experiments at decreasing values of $V_{in}$ to selectively actuate fewer parts of the system and test its performance and reliability at finer scales. We track the torsion frame first as it drives the ring gear and load gear, then after decoupling the torsion frame from the ring gear to eliminate mechanical feedback, and finally as the system manifests motion due to friction and wear leading to terminal failure.

### A. Normal Operation

$V_{in} = 15$ V results in sufficient motion of the torsion frame to engage the ratchet pawls and rotate the ring gear and load gear. The torsion frame reaches the same *on* position due to a stiff stop, allowing verification of measurement uncertainty as the only source of variability for this *on* position (Fig. 2(b)). Interestingly, the *off* positions vary significantly over a range of 2 $\mu$m in the x direction, resulting in a broad and asymmetric distribution of $\Delta\theta$ (Fig. 2(c)). In all scatter plots, a lone vertical gray bar indicates measurement uncertainty in the y direction, while uncertainty in the x direction is smaller than the data markers. In all time series, uncertainties are comparable in size or smaller than the data markers, and we reduce the data density for clarity. The variation of *off* position is too large to be a characteristic of the input signal [4], and could be a result of the mechanisms that couple the torsion frame within the system.

### B. Feedback Effects

$V_{in} = 13$ V reduces the motion of the torsion frame so as to not engage the ratchet pawls and rotate the ring gear. The torsion frame motion becomes much less variable, to within approximately 300 nm in the x direction (Fig. 3). This indicates that mechanical feedback from rotation of the ring gear and load gear causes much of the variation of *off* position during normal operation (Fig. 2(a)). The ring gear connects to the substrate by guides (Fig.1(a)), which have clearances that allow the position of the ring gear to vary and transfer back to the torsion frame. While variation of the torsion frame position reduces upon decoupling from the motion of the ring gear, another behavior becomes apparent as a clustering at specific positions, causing transient decreases in the magnitude of $\Delta\theta$ (Fig. 3(c)). Some of the remaining variability may be due to unintentional motion of the measurement system [4], and we interpret the clustering of positions as sticking due to friction.

### C. Stiction Effects

$V_{in} = 10$ V results in a dramatic increase of sticking behavior (Fig. 4(a) and (b)), producing a complex distribution of $\Delta\theta$ (Fig. 4(c)). Unlike at higher $V_{in}$, the decrease in driving torque allows static friction to arrest the motion of the torsion frame at varying positions near the boundaries of its range. Wear is another potential factor contributing to sticking. Our test system includes a fluoropolymer coating [16], and our data suggest that this coating wears during operation, increasing friction [17]. Stiction likely results from friction between the torsion frame and the ring gear. A previous study of the reliability of this type of system observed, through much coarser and slower measurements, a failure mode caused by sticking of the ratchet pawls to the ring gear [14]. Such motion differs from the motion of the ring gear that occurs when the ratchet pawls engage, and could explain the variation in position. Autocorrelation of both *off* and *on* $\theta$ reveals a negative correlation between consecutive motion cycles, and cross-correlation of *off* and *on* $\theta$ reveals a negative correlation between each *on* $\theta$ and the consecutive *off* $\theta$ (not shown). These results indicate the dynamic effects of energy storage in the torsion beams and relief springs, and energy dissipation by friction. Moreover, correlation at longer lag times indicates additional periodicity of the motion, which we will study in the future.

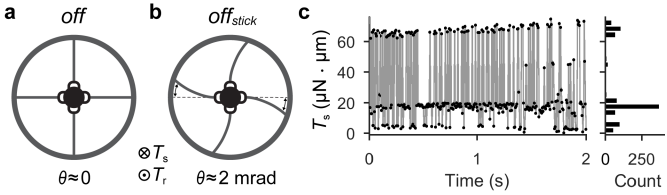

Fig. 5. Stiction torque. (a) Schematic showing the torsion frame and torsion beams in an *off* position at $\theta \approx 0$. (b) Schematic showing the torsion frame and torsion beams in an *off* position at $\theta \approx 2$ mrad. (c) Time series and histogram showing the stiction torque acting on the torsion frame at each *off* position.

*D. Stiction Torque*

We estimate the magnitude of the stiction torque $T_s$ from a torque balance, a nominal model relationship between the torque of the four torsion beams and relief springs as a function of $\theta$ [13], and the sticking points in Fig. 4. Three torques act in the motion plane on the torsion frame – the driving torque $T_d$ from the electrostatic comb array rotating it to an *on* position, the restoring torque $T_r$ from the torsion beams and relief springs rotating it back to an *off* position, and $T_s$ opposing all motion. In the *off* positions, $V_{in} \approx 0$, so that $T_d \approx 0$ and $T_r = T_s$. Neglecting any deviation of the four torsion beams and relief springs from the model, such as from variation of material and dimensional properties within the engineering tolerance of the microfabrication process, the restoring torque in $\mu$N·$\mu$m is $T_r = 4(2.60\theta + 3.59 \times 10^{-3}\theta^2 - 8.77 \times 10^{-6}\theta^3)$ [13], where $\theta$ is in radians. We measure $\theta$ with respect to the orientation of the torsion frame for $T_d = T_r = T_s = 0$, which we approximate as its most counterclockwise orientation in the time series. We calculate the value of $T_r = T_s$ at each *off* position (Fig. 5). Our $\Delta\theta$ uncertainty of 33 $\mu$rad gives a mean uncertainty of torque measurements of 1.1 $\mu$N·$\mu$m. Increases in the largest values of $T_s$ and the emergence of sticking corresponding to $T_s \approx 45$ $\mu$N·$\mu$m indicate changing surface properties. Application of our method to measure MEMS designed for tribology tests [17] can provide further insight into the relevant effects that produce these complex dynamics.

## IV. Conclusion

Following these experiments, stiction forces in the *off* positions became large enough to require an order of magnitude increase of $V_{in}$ to produce motion. Under these conditions the comb fingers locked irreversibly in direct contact (Fig. 1(a)), causing terminal failure of the system.

Ongoing efforts to improve the performance and reliability of kinematic assemblies in MEMS would benefit generally from better measurements of motion. This is a challenge, as the limiting motions can occur in multiple degrees of freedom, and can be small, fast, and transient. By advancing and applying our particle tracking method, we have studied coupling interactions that result in unintentional motion of parts through mechanical feedback, and the effects of friction and wear between parts in sliding contact leading to system failure. These effects can be critical to MEMS operation but have not been previously measured *in operando* at nanometer, microradian, and millisecond scales. Our measurements point the way toward understanding and improving the dynamic behavior of MEMS.